\documentclass[onecolumn,showpacs,amsmath,amssymb]{revtex4}
\usepackage[dvips]{graphicx}

\usepackage{dcolumn}
\usepackage{bm}

\begin{document}

\title{Compact optical grating compressor.}


\author{Vladyslav V. Ivanov}
\email{vladivanov78@gmail.com}

\affiliation{Physical Sciences Inc. 20 New England Bus Center Dr, Andover, MA 01810 USA}




\begin{abstract}
A novel design of a grating-based optical pulse compressor is proposed. The proposed compressor provides a large group delay dispersion while keeping the compressor linear size small. The design of the proposed compressor is based on a traditional Treacy compressor with a straightforward modification of inserting two lenses between the compressor's gratings. This simple alternation aims to substantially increase group delay dispersion of the compressor or alternatively to decrease the compressor size while maintaining its group delay dispersion. A theoretical description of the enhanced compressor has been developed in the paraxial approximation.  A detailed numerical model has been built to calculate the compressor parameters more accurately. These theoretical studies have revealed that the enhanced optical  compressor provides a significant increase in the group delay dispersion compared to a standard Treacy compressor.
\end{abstract}

\maketitle
\section{Introduction}
In laser amplifiers for ultra-short pulses, the peak optical powers become very high, so nonlinear pulse distortion or destruction of some optical components might occur. Chirped pulse amplification (CPA) is a common technique to circumvent this problem. CPA for lasers was invented in the mid-1980s, \cite{Strickland85} and awarded the  Nobel Prize in Physics in 2018.
CPA is routinely used in high peak power lasers. Ultrafast high power lasers have numerous applications such as high-precision material machining including drilling, cutting, and surface processing \cite{Orazi21, Zhang19, Lei20, Theodosiou19, Xu19}, medical applications, in particular,  ophthalmic surgery  \cite{Lubatschowski03, Li22}, military applications like  countermeasures or direct energy weapon \cite{LaserWeapon21, SBIR2017, SBIR2020}.

Peak powers of 10 TW and above have been reported \cite{Tavella07, Rouyers93}.  Currently, laser systems delivering peak power of Terrawatts-level can operate in the laboratory environment only due to their size, weight, and frequent need for maintenance \cite{Matsuoka06, Kessler14, Ekspla, Lightcon}.
 However, high peak power laser systems would be interested in a number of applications outside laboratory environments such as manufacturing and defense \cite{Ekspla, Breitkopf14, Chvykov17, SBIR2020, devcom2021}.

In the CPA technique, low-power short pulses are passed through a stretcher i. e. optical element with a positive spectral dispersion so that the pulses are temporally stretched prior to amplification. The stretched pulses can be safely amplified without damaging the amplifier material while the peak power significantly reduced. After the amplification, the original pulse width is recovered by passing the pulses through the compressor, which cancels the positive dispersion of the stretcher by providing the dispersion equal in amplitude but opposite in sign. The pulse compressor is based on a concept introduced by Treacy in 1969. The Teacy$'$s concept relies on the grating pair
that provide large values of the negative dispersion.

Currently, available peak powers are limited by the optical damage of the amplifier material. The optical damage can be alleviated by stretching the pulses to a longer duration.
Significant difficulties are posed by the available diffraction gratings.
First, it is difficult to fabricate gratings with large sizes that are uniformly ruled over large areas.
This imposes a limit on the peak powers and pulse duration achievable through the use of conventional compressors. Second, large dispersion compressors imply a large distance between the compressor's elements. This limits the size of the laser system and makes their use unpractical outside of labs, especially on moving vehicles.

The linear size of the optical compressor can be significantly reduced by folding the beam path using retrospective prisms \cite{Lai93} or by integrating diffraction
gratings onto opposite surfaces of a solid fused silica block  \cite{Yang16}.
While Yang and Towe \cite{Yang16} design offers an elegant idea that may be difficult to manufacture. Further, their design implies propagation of the laser pulse inside of the media for a significant distance that might cause spectral distortions or optical damage.
The design proposed by Lai \cite{Lai93} is easily achievable, and can be further improved using grism \cite{Chauhan10}.  However, the design presented by Lai doesn't address the problem of a long optical path. In a way the design described in \cite{Lai93}  trades a linear size of compressor to several reflective optical elements, thus such an optical compressor remains bulky and sensitive to vibrations.

The present work focuses on reducing the required optical path by enhancing the dispersion of the compressor.
In this work, group delay dispersion of the compressor is substantially enhanced by two cylindrical lenses placed between the compressor$'$s gratings. The theoretical study below demonstrates a notable advantage of such a compressor compared to the more standard design.

\section{Description and Theory of the Treacy Grating Compressor}

The basic sketch of the traditional Treacy grating compressor \cite{Treacy69} is shown in Fig.1(a).
It comprises two parallel diffraction gratings, with their working surfaces facing one another and their lines parallel to each other.
When a beam is an incident on the first grating of the compressor (A), its spectral components of the incoming beam are diffracted at different angles.
In a sense, the grating acts as a convex mirror or a negative lens in the case of a transmission grating.  After the reflection, the different spectral components of the beam become spatially separated.
After the reflection from the second grating (B), the beam becomes again collimated.
 At the output of the grating pair, the beam is spatially incoherent.
 This can be solved by a 2nd identical gratings pair or more practically by retro-reflecting the light back into original the grating pair by a mirror (C).
Additionally, that generates double the amount of negative dispersion. Ultimately the different spectral components have different, frequency-dependent optical paths. This path difference creates a negative dispersion, which is needed for the re-compression of previously stretched pulses.
The group delay dispersion (GDD) of the Treacy compressor is written in Eq.1.

\begin{figure}[h!]
\centering\includegraphics[width=7cm]{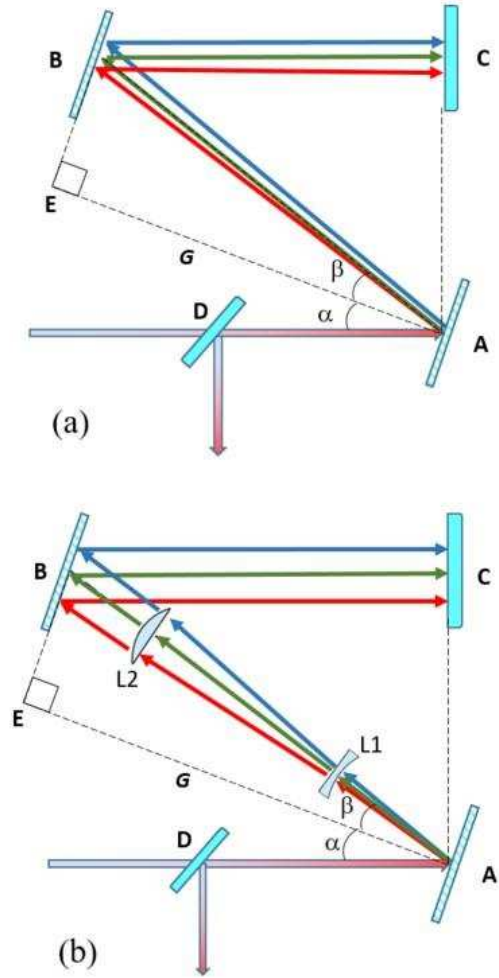}
\caption{ (a) the Treacy grating compressor,
(b) the enhanced grating compressor.}
\end{figure}

\begin{equation}
\varphi'' = -\frac{G  \lambda^{3}}{2 \pi c^2 d^{2} \cos^{3}\beta},
\end{equation}
here $G$ is the slant (perpendicular) distance between the gratings, $\beta$ is the reflection angle from the grating, $d$ is the grating constant, and $c$ is the speed of light.
The reflection angle $\beta$ can be found in Eq. 2 as

\begin{equation}
cos\beta=\sqrt{1-(\sin\alpha-\lambda/d)^{2}},
\end{equation}
here $\alpha$ is the incident angle of the incoming beam to the 1st grating.

The dispersion provided by the traditional Treacy compressor $\varphi''$ is a function of gratings period $d$, the beam incidence angle $\alpha$, and distance in the compressor $G$ .
Unfortunately, increasing the compressor's dispersion using the first two parameters is problematic. The gratings period can not be arbitrarily small because of manufacturing constraints, and the incidence angle is usually chosen to maximize power efficiency. Larger negative dispersion in grating compressors can be achieved by increasing the distance between the gratings. For moderate pulse duration with several hundreds of picoseconds, a typical compressor length is in order of tenth centimeters. However, if one has to stretch the pulse to the duration of 1 nsec or above, propagation distances beyond a meter are required. As a simple yet practical example, we consider a pulse with a Fourier limited width of 500 fsec (FWHM), which corresponds to the wavelength spread of $\Delta\lambda$ =3.1 nm (FWHM) at the central wavelength of 1030 nm. The wavelength of 1030 nm is chosen because it is the optimal wavelength of Yb:YAG lasers that are a popular choice for high-power ultrafast lasers.
 The pulse is stretched to a temporal width of 1 nsec, using chirped fiber Bragg grating (CFBG) with the second-order group velocity dispersion of 180 $ps^2$.
The non-linear effects are neglected assuming strong pulse stretching keeps the B-integral values small.
In order to compress the pulse, the grating compressor has to have the dispersion of the opposite sign and equal amplitude. If for example, one uses the commercially available transmission grating $T-1702-1030$ from Lightsmyth, with a groove density 1702.13 mm$^{-1}$, and the incidence angle of 61.2$^{\circ}$, the required grating separation $G$ would be 1.8 meters in double-pass grating pair compressor in Treacy configuration.
This is an impractically long optical path, that makes the compressor, and hence the laser system, heavy, bulky, sensitive to vibrations, and overall unusable.

\section{Description of the Compact Grating Compressor}

The path differences for different spectral components can be significantly increased
 by modifying the traditional Treacy compressor design as shown in Fig. 1(b).
The spectral chirp of the beam reflected by 1st grating (A) can be further enhanced by a negative cylindrical lens (L1) placed between 1st and 2nd gratings.
The L1 is placed at a position where the beam is already spatially chirped, thus, it selectively addresses different spectral components of the beam, enhancing the propagation path for 'red', while decreasing it for the 'blue' spectral components.
Further, the effect of the lens L1 on the beam divergence is compensated by a positive lens (L2) that recovers the beam divergence to its value prior to the lens L1. The lenses L1 and L2 are positioned in the manner that the beam peak intensity and hence the central wavelength passes through the center of both of them.

Analytical estimations have been performed assuming a small angular divergence of the beam reflected from the 1st grating, i.e. in paraxial approximation. Fig. 2 provides a more detailed scheme of the proposed grating compressor. The transmission gratings are used for a practical example because of their high diffraction efficiencies.
Propagation of two spectral components is considered and noted as a  trial and a reference beam. The reference beam propagates through the center of the lenses L1 and L2, hence its propagation path is not altered. For the practical reason of avoiding optical aberrations, the reference beam should correspond to a  spectral component of a central wavelength of the compressed beam.
The trial spectral components propagate off the center of L1 and L2 hence its propagation direction changes so its optical path.
 In Fig. 2,  $x_{1}$, $x_{2}$ and $x_{3}$ are the distances between the 1st diffraction grating and the lens L1, lenses L1 and L2, and the lens L2 and 2nd diffracting grating correspondingly ( $x_{1}=AH_{1}$, $x_{2}=H_{1}F_{1}$ and $x_{3}=F_{1}B$ ). $G_{1}$, $G_{2}$ and $G_{3}$ are the projection of the distances $x_{1}$, $x_{2}$ and $x_{3}$ on the axis perpendicular to the 1st and 2nd gratings. The three projections combined are equal to the total slant distance between gratings, $G_{1}+G_{2}+G_{3}=G$, where $G=DB_{1}$.

 \begin{figure}[h!]
\centering\includegraphics[width=12cm]{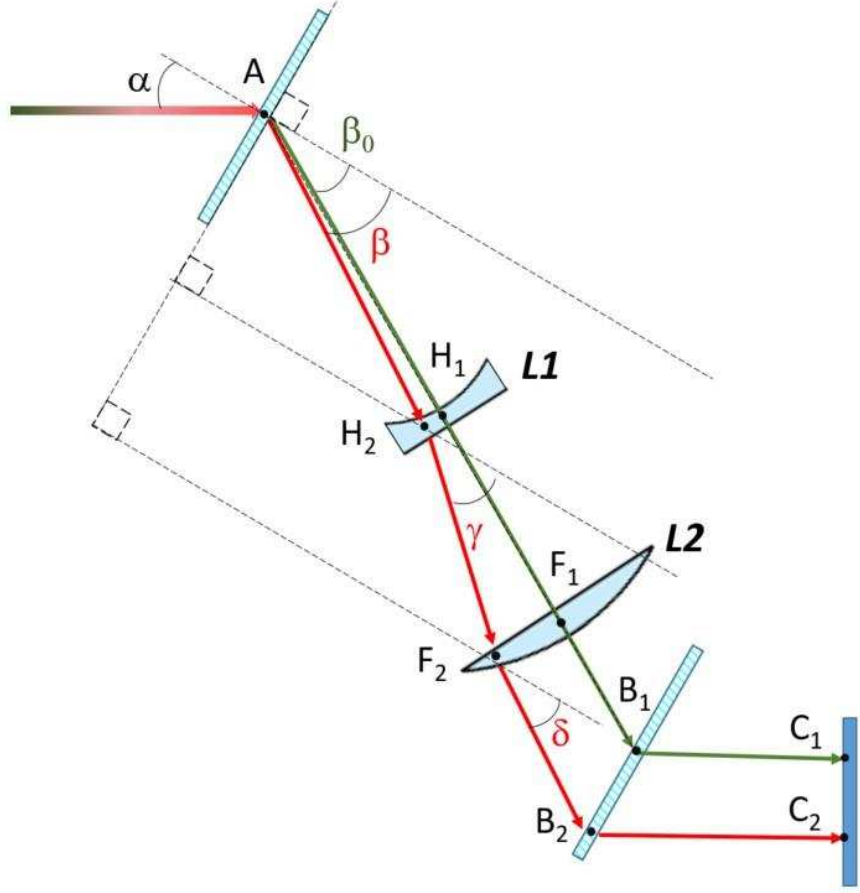}
\caption{Propagation of the reference and trial beams. The reference spectral component reflects from the 1st grating at the angle $\beta_{0}$ and propagates through the points $AH_{1}F_{1}B_{1}C_{1}$ through the center of both lenses.
The trial spectral component reflects from the 1st grating at the angle $\beta$  propagates through the points $AH_{2}F_{2}B_{2}C_{2}$. }
\end{figure}

The original Treacy paper \cite{Treacy69} describes a straightforward and intuitive path for calculating the GDD that has been directly applied to the enhance compressor. As it was shown there, the GDD can be written as in Eq. 3.

\begin{equation}
\varphi''  = \frac{d^{2}\varphi}{d\omega^{2}} = \frac{1}{c} \frac{dp}{d\beta} \frac{d\beta}{d\omega},
\end{equation}

 where $p$ is the optical path, and $\omega$ is the frequency. First term $d\beta/d\omega$ read as Eq. 4.

\begin{equation}
 \frac{d\beta}{d\omega} = \frac{2\pi c }{ \omega^{2} d \cos\beta}.
\end{equation}

To find $dp/d \beta$ the optical path $AB_{2}C_{2}$ was spitted in three parts: before L1 ($G_{1}$), in between L1 and L2 ($G_{2}$), and after L2 ($G_{}3$). Assuming that the L2 recovered the initial propagation angle before L1 i.e. $\delta = \beta$  the problem is identical to the standard Treacy compressor for the first and third parts.
Their pass $p$ can written as $p=(G_{1}+G_{3}) \cdot [1+cos(\alpha+\beta)]/cos(\beta)$, one straightforwardly gets $\frac{dp_{1}}{d\beta}$  as shown in Eq. 5.

\begin{equation}
\frac{dp_{1}}{d\beta}  = - \frac{G_{1} \lambda}{d \cos^{2}\beta}.
\end{equation}

The 2nd part of the optical path (i.e. after L1, before L2), is affected by the negative length L1. The angle between the grating and a spectral component propagation changes, the new angle is noted as $\gamma$.  The variation of the optical path as a function of the angles $\gamma$, $\beta$, and hence the wavelength $\lambda$ leads to modification of the GDD of the compressor.
Because the optical path after the lens L1 doesn't depend directly on the angle $\beta$, but only indirectly through the angle $\gamma$, which is the function of $\beta$. One can write Eq. 6.

\begin{equation}
\frac{dp_{2}}{d\beta}  = \frac{dp_{2}}{d\gamma} \frac{d\gamma}{d\beta},
\end{equation}
here $p_{2}$ is the optical path between the lenses L1 and L2 ($H_{2}F_{2}$).

The calculation of the derivative $\frac{dp_{2}}{d\gamma} $   is analogous to the calculations of $\frac{dp_{2}}{d\beta}$ .

The offset of the trial spectral component on the lens L1 is $H_{1}H_{2}=x_{1}\tan(\beta-\beta_{0})$, where the $\beta$ and $\beta_{0}$ are the reflection angles from the 1st grating for the trial and reference beams correspondingly. Using the Ray transfer matrix analysis (ABCD matrix formalism) \cite{ABCD}   the expression for the angle $\gamma$ in Eq. 7 has been obtained,
where the $\gamma$ is the angle between the spectral component and the gratings.

\begin{equation}
\gamma = \beta +  \frac{G_{1}\tan(\beta-\beta_{0})}{f_{1} \cos\beta_{0}},
\end{equation}
where $f_{1}$ is the focal distance of the lens L1. Please, note that the angle $\beta$ is the function of the spectral component wavelength.
Eq. 7 is an approximation assuming small angles between spectral components and the lens axis.

Assembling the equation (5-7) and performing derivative $\frac{d\gamma}{d\beta} $
the expression for the derivative    $\frac{dp_{2}}{d\beta}$ is derived.

\begin{equation}
\frac{dp_{2}}{d\beta}  = - \frac{G_{2} (\sin\alpha-\sin\gamma)}{ \cos^{2}\gamma} \left(1+\frac{G_{1} \sec^{2}(\beta-\beta_{0})}{f_{1} \cos\beta_{0}} \right),
\end{equation}

Overall expression for the GDD of the modified compressor is written in Eq. 9.

\begin{equation}
\varphi'' = -\frac{\lambda^2} {2 \pi  c^2 d  \cos\beta}\left[\frac{(G_{1}+G_{3})\lambda }{d \cos^{2}\beta} + \frac{G_{2} (\sin\alpha-\sin\gamma) }{\cos^{2}\gamma}\left(1+\frac{G_{1} \sec^{2}(\beta-\beta_{0})}{f_{1} \cos\beta_{0}} \right)\right],
\end{equation}

here the angle $\gamma$ is the function from angles $\beta$ and  $\beta_{0}$  as shown in Eq. 7, while the angles  $\beta$ and  $\beta_{0}$  are the function of the incidence angle $\alpha$ (Eq. 2). Eq. 9 provides a theoretical description of GDD based on input parameters, such as the incidence angle, the wavelength, the grating period, geometrical distances, and the focal distance of the lens L1. Note, that Eq. 9 presents single-pass  GDD, in the case of the 2nd pass provided by retro-reflection the results should be multiplied by 2.
The expression above is rather complicated, but one can get some intuitive understanding assuming $\beta \approx \beta_{0}$. In this case, the optical path isn't altered significantly by the lens L1, however, this doesn't mean that its derivative vs. $\omega$ isn't altered significantly and hence, the  GDD. If $\beta \approx \beta_{0}$ the second term in Eq. 7 is small compared to the first one, in this case, $\beta \approx \gamma$.
Based on these assumptions Eq. 9  can be simplified to Eq. 10.

\begin{equation}
\varphi'' = -\frac{\lambda^{3}}{2 \pi c^2 d^{2} \cos^{3}\beta}\left[G_{1}+G_{3}+G_{2} \left(1+\frac{G_{1} }{f_{1} \cos\beta} \right)\right].
\end{equation}

Based on Eq. 10 one can make some preliminary conclusions and estimate the parameters giving the highest GDD. The $G_{3}$ should be kept as small as physically possible since it doesn't contribute to the enhancement of GDD while placing L1 in the middle between the 1st grating and L2 ($G_{1}=G_{2}$) provides the maximum dispersion. This is a compromise between the necessity to have the beam spatially chirped in the direction perpendicular to the beam propagation and enough propagation length to benefit from this enhanced chirp. The focal length $f_{1}$ should be as short as possible and limited mostly by manufacturing considerations and geometrical constraints. The dependence of the GDD on other parameters such $d$, $\lambda$, and $\beta$ is similar to the standard Treacy compressor.

I made estimations of the performance of the enhanced compressor based on Eq. 10.
The considered parameters are identical to the parameters in the Section 2: the $\lambda$=1030 nm,  the wavelength spread of $\Delta\lambda$ =3.1 nm (FWHM), the grating period $1/d=1702$ per mm, $\alpha=61.2^{\circ}$. The angle spread of the reflected beam is $  0.6 ^{\circ}$.
I assume the L1 is positioned in the middle between the gratings, and the distance the G3 is small
(i.e. $ G_{3} \ll G_{1}, G_{2} $).
Assuming the slant distance between gratings $G$ equals $1$ m and the focal distance $f_{1}$ of L1 equals $-0.1$ m, one calculates that the pair of lenses L1 and L2 lead to a significant increase in the GDD of the compressor by a factor of 6.2. Such an increase in the GDD allows the achievement of higher peak powers by alleviating the risk of optical damage.

\section{Numerical simulations}

To go beyond the assumptions of the previous section the numerical model was developed based on MATLAB software. The model calculates the optical path depending on the wavelength, then
the model calculated changes of the optical path versus small variation of the angle $\beta$,  basically numerically calculating the derivative
$dp/d\beta$. Such calculations directly provide the dependence of the GDD versus wavelength. These calculations are performed for 500 wavelengths in the range between 1025 and 1035 nm, thus yielding a GDD versus the wavelength curve that can be compared with a GDD versus the wavelength curve for a Treacy compressor with similar parameters.

The key task addressed by the numerical model is the calculation of the angles  $\gamma$ and $\delta$ beyond the ABCD matrix expression used for Eq. 7.
In the general case, the angles $\gamma$ and $\delta$ depend on the specific shape of the lens. I assume the pair of fused-silica plano-concave and plano-convex lenses were employed for L1 and L2. The lenses L1 and L2 are positioned in the way that the 1030 nm spectral component passes through the center of both lenses at the right incidence angle thus experiencing no angle change.
After passing through the plano-concave lens a beam propagation angle of $\varphi_{in}$ changes to an angle $\varphi_{out}$ that can be written as shown in Eq. 11.

\begin{equation}
\varphi_{out} = \arcsin\left\{  n\times\sin\left\langle \arcsin\left[
\frac{\varphi_{in}+\arcsin(y/R)}{n}
\right]-\arcsin(y/R) \right\rangle \right\}
\end{equation}
here $R$ is the curvature radius of the lens,  $n$ is the refractive index of the lens material and $y$  is the displacement from the entrance of the lens.

Further, the numerical model addresses the finite (non-zero ) thickness of the lenses L1 and L2.

\begin{figure}[h!]
\centering\includegraphics[width=10cm]{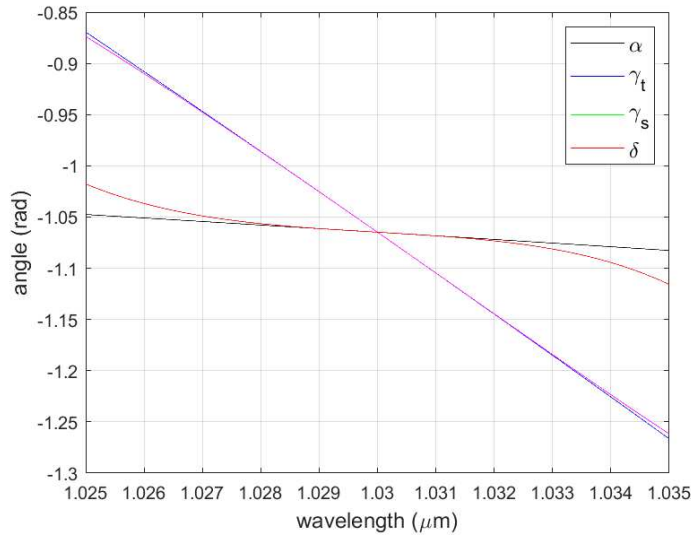}
\caption{The angles for the spectral components are presented as a function of the wavelength. The black line shows the angle $\beta$ after the reflection from the 1st grating. The magenta  and blue curves show the angle $\gamma$ after the lens L1 and were calculated based on Eq. 8 and numerical simulations correspondingly.
The red curve shows the angle $\delta$  after the lens L2 based on the numerical model.
 }
\end{figure}

First, I calculate the angles of $\beta$, $\gamma$, and $\delta$  depending on the wavelength shown in Fig. 3. The angles  $\gamma$ are calculated based on the theory Eq. 7 and the numerical model. All curves are calculated for same parameters as before:  $\lambda$=1030 nm,  $\Delta\lambda$ =3.1 nm,
$1/d$=1702 per mm,  $\alpha=61.2^{\circ}$, G=1 m, $G_{1}$=0.5 m, $G_{2}$=0.4 m and $G_{3}$=0.1 m. The L1 and L2 lenses are centered on a 1030 nm spectral component and have the focal distances of $f_{1}=$-0.1 and $f_{2}=$0.5 m correspondingly.
As one can see Eq. 7 provides a good approximation in the wavelength range close to the central $\lambda_{0}=1030$ nm because the spectral components of the beam are passing near the center of lenses L1 and L2. The output angle $\delta$ theoretically should well coincide with the original angle $\beta$ assuming that the focal distance $f_{2}$ is chosen correctly, however one can see noticeable deviation at the wavelengths 1025 nm and 1035 nm. Such deviations occur because Eq. 7 is not valid for a significant distance between a spectral component and the lens center. The beam diameter increased substantially more before L2 compare to L2 the paraxial approximation is rougher for the angles $\delta$ compared to the angles $\gamma$.

\begin{figure}[h!]
\centering\includegraphics[width=10cm]{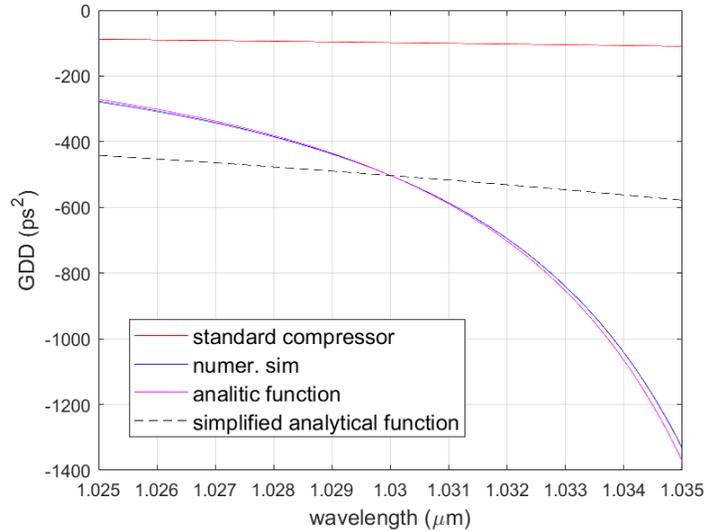}
\caption{GDD is presented as a function of the wavelength for a number of scenarios. The red line is GDD for the traditional Treacy compressor for the same grating parameters and gratings separation ($G$, $\alpha$, and $1/d$ ). All other plotted curves are calculated for the enhanced compressor. The black, dashed curve is the calculation based on simplified Eq. 10. and the magenta curve is calculated based on Eq. 9. Finally the blue curve is a result of the numerical simulations described above.
 }
\end{figure}

I calculate the GDD as a function of the wavelength using the numerical model and compare it with the traditional Treacy compressor as well as analytical
solutions full Eq.9 and simplified  Eq.10. The parameters of standard Treacy and enhanced compressor are identical aside from the lenses L1 and L2 used in the enhance compressor.
The results are presented in Fig. 4. Figure 4 is the main output of this paper.
First, one can see the significant enhancement of the GDD compared to a standard Treacy compressor (red curve). Overall the approximation based on Eq. 9 (blue curve) does a decent job, while further simplified Eq. 10 (black dashed curve) predicts expected GDD only at the central wavelength and its vicinity. The deviation between the numerical model and theoretical expression of Eq. 9 is within 5 $\%$ in a chosen wavelength range. Near the central wavelength of 1030 nm, the deviation is negligible. For these conservative, realistic parameters we calculate the lenses L1 and L2 increase the GDD of the compressor by more than a factor of 5 at the wavelength of 1030 nm.  The discrepancy between the enhancement of the factor of 5 in Figure 5 and the enhancement of the factor of 6.2 roughly estimated at the end of the previous section is caused mostly because of setting $G_{3}$ to 0 for the rough estimations.

Increased curvature of the GDD vs. wavelength line indicates higher values of higher order desertions that have to be compensated by CFBG.

\begin{figure}[h!]
\centering\includegraphics[width=10cm]{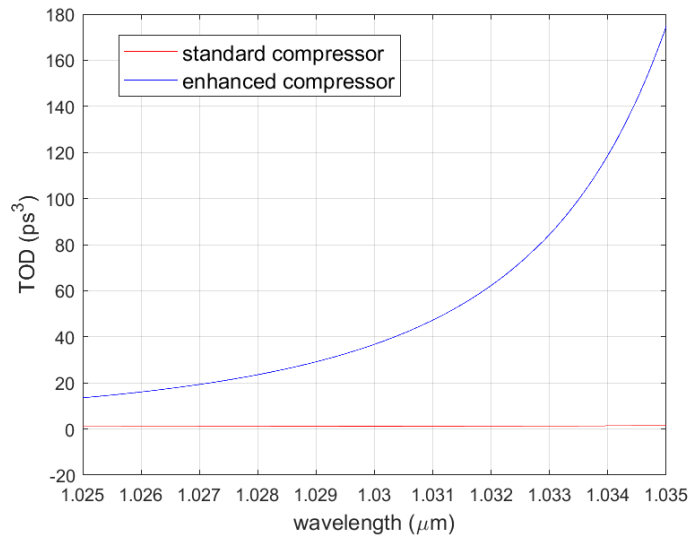}
\caption{TODs as a function of the wavelength are presented for a traditional Treacy, and enhanced compressor (the same parameters ($G$, $\alpha$ and $1/d$ for both compressors).
 }
\end{figure}

The third-order dispersion (TOD) of the enhanced compressor becomes important for broad-band pulses. Theoretically, it can be pre-compensated by CFBG, however, in order to do so the TOD value of the later stages needs to be known.
Although it is possible to calculate TOD by taking the derivative of Eq. 8 over $\lambda$ analytically
the usefulness of this step is questionable because of the length and complexity of the result.
Instead, the TOD can be calculated straightforwardly using a numerical model. In Fig. 5 we show TOD as a function of the wavelength for a standard Treacy compressor and for an enhanced compressor with parameters the same as above. We observe a TOD increase of a factor of 40 at the wavelength of 1030 nm.

Such large TOD has to be pre-compensated by the design of  CFBG or by employing tunable FBG \cite{Min18}.

\section{Conclusion}
I have proposed an enhanced optical grating compressor that is suitable for compressing strongly stretched laser pulses.
This novel compressor employs cylindrical lenses to enhance the spectral chirp of the beam inside of the compressor that
allows achievement of significantly larger GDD compare to currently used Treacy compressors.
This drastically reduces the required linear distances; which makes the compressor compact and more suitable for out-of-the-lab applications. The enhanced compressor allows to greatly reduce the size of the high-peak-power lasers, thus making them more applicable outside of scientific laboratories, in particular, on moving vehicles.

\section{Acknowledgements}
The author thanks Kevin Wall and Bhabana Pati for their careful reading of the manuscript and insightful suggestions.




\begin{thebibliography}{1}
\newcommand{\enquote}[1]{``#1''}


\bibitem{Strickland85}
Donna Strickl and Gerard Mourou
   \enquote{Compression of amplified chirped optical pulses,}
Optics Communications \textbf{56(3)}, 219-221 (1985).

\bibitem{Orazi21}
L. Orazi, L. Romoli, M. Schmidt and L. Li,
   \enquote{Ultrafast laser manufacturing: from physics to industrial applications,}
CIRP Annals \textbf{70(2)}, 543-566 (2021).

\bibitem{Zhang19}
Jie Zhang, Sha Tao, Brian Wang, and Jay Zhao
   \enquote{Development of industrial scale laser micro-processing solution for mobile devices,}
SPIE LASE \textbf{10906}, 109061A  (2019).

\bibitem{Lei20}
Shuting Lei, Xin Zhao, Xiaoming Yu, Anming Hu, Sinisa Vukelic, Martin B. G. Jun, Hang-Eun Joe, Y. Lawrence Yao, Yung C. Shin,
   \enquote{Ultrafast Laser Applications in Manufacturing Processes: A State of the Art Review,}
Journal of Manufacturing Science and Engineering \textbf{142(3)}, 1-43 (2020).

\bibitem{Theodosiou19}
Antreas Theodosiou, Rui Min, Arnaldo G. Leal-Junior, Andreas Ioannou, Anselmo Frizera, Maria Jose Pontes, Carlos Marques, and Kyriacos Kalli,
   \enquote{Long period grating in a multimode cyclic transparent optical polymer fiber inscribed using a femtosecond laser,}
Optics Letters \textbf{44(21)},  5346-5349 (2019).

\bibitem{Xu19}
K. Xu, Y. Chen, T. A. Okhai, and L. W. Snyman,
   \enquote{Micro optical sensors based on avalanching silicon light-emitting devices monolithically integrated on chips,}
Optical Materials Express \textbf{9(10)}, 3985-3997 (2019).

\bibitem{Lubatschowski03}
Holger Lubatschowski, Alexander Heisterkamp, Fabian Will, Ajoy I. Singh, Jesper Serbin, Andreas Ostendorf,
Omid Kermani, Ralf Heermann, Herbert Welling, and Wolfgang Ertmer,
   \enquote{Medical applications for ultrafast laser pulses,}
Proceedings of SPIE - The International Society for Optical Engineering \textbf{50(50)},  (2003).

\bibitem{Li22}
C. L. Li, C. J. Fisher, R. Burke, and S. Andersson-Engels,
   \enquote{Orthopedics-Related Applications of Ultrafast Laser and Its Recent Advances,}
Appl. Sci. \textbf{12}, 3957 (2022).

\bibitem{LaserWeapon21}
www.militaryaerospace.com/power/article/14206874/laser-weapons-ultrashortpulse-fiber

\bibitem{SBIR2017}
www.sbir.gov/node/1207829

\bibitem{SBIR2020}
www.sbir.gov/node/1654485

\bibitem{Tavella07}
Franz Tavella, Yutaka Nomura, Laszlo Veisz, Vladimir Pervak, Andrius Marcinkevičius, and Ferenc Krausz,
   \enquote{Dispersion management for a sub-10-fs, 10 TW optical parametric chirped-pulse amplifier,}
Opt. Lett. \textbf{32(15)}, 2227 (2007).

\bibitem{Rouyers93}
C. Rouyer, É. Mazataud, I. Allais, A. Pierre, S. Seznec, C. Sauteret, G. Mourou, and A. Migus,
   \enquote{Generation of 50-TW femtosecond pulses in a Ti:sapphire/Nd:glass chain,}
Opt. Lett. \textbf{18}, 214 (1993).

\bibitem{Matsuoka06}
Shinichi Matsuoka, Takehiro Yoshii, Masatoshi Sato, Fumihiko Nakano, Yoshimori Tamaoki, You Wang, Yoshinori Kato, Koichi Iyama, Minoru Nishihata, Hirofumi Kan, and Sadao Nakai,
   \enquote{1TW, 10 Hz and 0.1TW, 1 kHz All-Solid-State Femtosecond Lasers,}
The Review of Laser Engineering \textbf{34(9)}, 214 (2006).

\bibitem{Kessler14}
Alexander Kessler, Marco Hornung, Sebastian Keppler, Frank Schorcht, Marco Hellwing, Hartmut Liebetrau, Jörg Körner, Alexander Sävert, Mathias Siebold, Matthias Schnepp, Joachim Hein, and Malte C. Kaluza,
   \enquote{16.6 J chirped femtosecond laser pulses from a diode-pumped Yb:CaF2 amplifier,}
Opt. Lett. \textbf{39}, 1333 (2014).

\bibitem{Ekspla}
www.ekspla.com/product/sylos-2a/

\bibitem{Lightcon}
www.lightcon.com/product/high-energy-opcpa/

\bibitem{Breitkopf14}
Sven Breitkopf, Tino Eidam, Arno Klenke, Lorenz von Grafenstein, Henning Carstens, Simon Holzberger, Ernst Fill, Thomas Schreiber, Ferenc Krausz, Andreas Tünnermann, Ioachim Pupeza, and Jens Limpert,
   \enquote{ A concept for multiterawatt fibre lasers based on coherent pulse stacking in passive cavities,}
Light: Science and Applications \textbf{3(e211)}, 1 (2014).

\bibitem{Chvykov17}
V. Chvykov
   \enquote{ New Generation of Ultra-High Peak and Average Power Laser Systems,}
Intechopen 70720 (2017).

\bibitem{devcom2021}
apps.dtic.mil/sti/pdfs/AD1125446.pdf

\bibitem{Lai93}
M. Lai, S. T. Lai, and C. Swinger,
   \enquote{Single-grating laser pulse stretcher and compressor,}
Applied Optics \textbf{33(30)}, 6985 (1993).

\bibitem{Yang16}
C. Yang, E. Towe,
   \enquote{Ultra-compact grating-based monolithic optical pulse compressor for laser amplifier systems,}
Journal of the Optical Society of America B \textbf{33(10)}, 2135-2143  (2016).

\bibitem{Chauhan10}
V. Chauhan, P. Bowlan, J. Cohen, and R. Trebino,
   \enquote{Single-diffractiongrating and grism pulse compressors,}
Journal of the Optical Society of America B \textbf{27(4)}, 619-624  (2010).

\bibitem{Treacy69}
E. B. Treacy,
   \enquote{Optical pulse compression with diffraction gratings,}
IEEE J. Quantum Electron. \textbf{5(9)}, 454 (1969).

\bibitem{ABCD}
www.rp-photonics.com/abcdmatrix.html

\bibitem{Min18}
Rui Min, Sanzhar Korganbayev, Carlo Molardi, Christian Broadway, Xuehao Hu, Christophe Caucheteur, Ole Bang, Paulo Antunes, Daniele Tosi, Carlos Marques, and Beatriz Ortega,
   \enquote{Largely tunable dispersion chirped polymer FBG,}
Opt. Lett. \textbf{43(20)}, 5106 (2018).

\end{thebibliography}


\end{document}